\documentclass[8.5pt,twoside,twocolumn]{article}
\usepackage[super,sort&compress,comma]{natbib}
\usepackage[version=4]{mhchem}
\usepackage{times,mathptmx}
\usepackage{sectsty}
\usepackage{balance}
\usepackage{graphicx} 
\usepackage{lastpage}
\usepackage[format=plain,justification=raggedright,singlelinecheck=false,font=small,labelfont=bf,labelsep=space]{caption}
\usepackage{fancyhdr}
\usepackage{graphicx}
\usepackage{dcolumn}
\usepackage{bm}
\usepackage{color}

\newcommand{\eqLabel}[1]{\label{eq:#1}}

\newcommand{\eref}[1]{Eq. (\ref{eq:#1})}

\newcommand{\bv}[1]{\mathbf{#1}}
\newcommand{\lr}[1]{\left( #1 \right)}
\newcommand{\LR}[1]{\left[ #1 \right]}
\newcommand{\qd}[1]{\left< #1 \right>}
\newcommand{\mat}[1]{ \underline{ \underline{ \bv{ #1 } } } }
\newcommand{\basis}[1]{\bv{\hat{#1}}}
\newcommand{\dv}[1]{#1'}
\usepackage{xcolor}

\usepackage{tikz}
\usetikzlibrary{calc}
\begin{document}
\thispagestyle{plain}
\fancypagestyle{plain}{
\fancyhead[L]{\includegraphics[height=8pt]{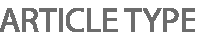}}
\fancyhead[C]{\hspace{-1cm}\includegraphics[height=20pt]{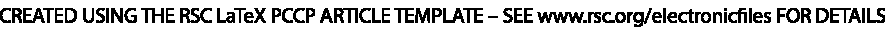}}
\fancyhead[R]{\includegraphics[height=10pt]{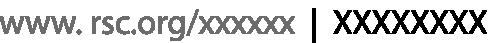}\vspace{-0.2cm}}
\renewcommand{\headrulewidth}{1pt}}
\renewcommand{\thefootnote}{\fnsymbol{footnote}}
\renewcommand\footnoterule{\vspace*{1pt}%
\hrule width 3.4in height 0.4pt \vspace*{5pt}}
\setcounter{secnumdepth}{5}

\makeatletter
\def\subsubsection{\@startsection{subsubsection}{3}{10pt}{-1.25ex plus -1ex minus -.1ex}{0ex plus 0ex}{\normalsize\bf}}
\def\paragraph{\@startsection{paragraph}{4}{10pt}{-1.25ex plus -1ex minus -.1ex}{0ex plus 0ex}{\normalsize\textit}}
\renewcommand\@biblabel[1]{#1}
\renewcommand\@makefntext[1]%
{\noindent\makebox[0pt][r]{\@thefnmark\,}#1}
\makeatother
\renewcommand{\figurename}{\small{Fig.}~}
\sectionfont{\large}
\subsectionfont{\normalsize}

\fancyfoot{}
\fancyfoot[LO,RE]{\vspace{-7pt}\includegraphics[height=9pt]{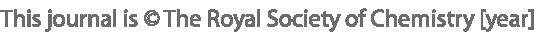}}
\fancyfoot[CO]{\vspace{-7.2pt}\hspace{12.2cm}\includegraphics{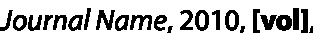}}
\fancyfoot[CE]{\vspace{-7.5pt}\hspace{-13.5cm}\includegraphics{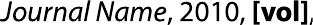}}
\fancyfoot[RO]{\footnotesize{\sffamily{1--\pageref{LastPage} ~\textbar  \hspace{2pt}\thepage}}}
\fancyfoot[LE]{\footnotesize{\sffamily{\thepage~\textbar\hspace{3.45cm} 1--\pageref{LastPage}}}}
\fancyhead{}
\renewcommand{\headrulewidth}{1pt}
\renewcommand{\footrulewidth}{1pt}
\setlength{\arrayrulewidth}{1pt}
\setlength{\columnsep}{6.5mm}
\setlength\bibsep{1pt}

\twocolumn[
  \begin{@twocolumnfalse}
\noindent\LARGE{\textbf{Laning, Thinning and
Thickening of Sheared Colloids
in a Two-dimensional Taylor-Couette Geometry.$^\dag$}}

\vspace{0.6cm}
\noindent\large{\textbf{Antonio Ortiz-Ambriz,\textit{$^{a}$} Sascha Gerloff,\textit{$^{b}$}
Sabine H. L. Klapp,\textit{$^{b}$}
Jordi Ort\'in,\textit{$^{ac}$}
and
Pietro Tierno\textit{$^{acd}$}}}\vspace{0.5cm}

\noindent\textit{\small{\textbf{Received Xth XXXXXXXXXX 20XX, Accepted Xth XXXXXXXXX 20XX\newline
First published on the web Xth XXXXXXXXXX 200X}}}

\noindent \textbf{\small{DOI: 10.1039/b000000x}}
\vspace{0.6cm}

\noindent \normalsize{
We investigate the dynamics and rheological properties of a circular colloidal cluster that is continuously sheared by magnetic and optical torques in a two-dimensional (2D) Taylor-Couette geometry.
By varying the two driving fields, we obtain the system flow diagram and report the velocity profiles along the colloidal structure.
We then use the inner magnetic trimer as a microrheometer, and observe continuous thinning of all particle layers followed by thickening of the third one above a threshold field.
Experimental data are supported by Brownian dynamics simulations.
Our approach gives a unique microscopic view on how the structure of strongly confined colloidal matter weakens or strengthens upon shear, envisioning the engineering of
rheological devices at the microscales.
}
\vspace{0.5cm}
 \end{@twocolumnfalse}
]

\section{Introduction}
\footnotetext{\dag~Electronic Supplementary Information (ESI) available:
Five .WMV videos showing the
dynamics of the colloidal clusters under combined magnetic and optical torques. See DOI: 10.1039/b000000x/}
\footnotetext{\textit{$^{a}$~Departament de F\'isica de la Mat\`eria Condensada, Universitat de Barcelona, Barcelona, Spain E-mail: ptierno@ub.edu}}
\footnotetext{\textit{$^{b}$~Institut f\"ur Theoretische Physik, Technische Universit\"at Berlin, Berlin, Germany}}
\footnotetext{\textit{$^{c}$~Universitat de Barcelona Institute of Complex Systems (UBICS), Universitat de Barcelona, Barcelona, Spain}}
\footnotetext{\textit{$^{d}$~Institut de Nanoci\`encia i Nanotecnologia, Universitat de Barcelona, 08028, Barcelona, Spain. }}
Understanding the dynamics of confined particulate systems under external deformations is relevant for many industrial and technological processes.~\cite{Mac94}
A classical yet versatile approach is based on the use of the Taylor Couette (TC) geometry, where complex fluids are confined and sheared between two coaxial cylinders.~\cite{Tay23,Far14}
The possibility to independently rotate these cylinders has made this geometry a powerful tool to investigate the emergence of centrifugal instabilities,
and how these flow perturbations lead to turbulence in a wide variety of soft matter systems, including foams,~\cite{Deb01,Lau04} granular
materials,~\cite{How99,Los00,Red11} micellar~\cite{Lop04} or polymeric solutions.~\cite{Gro96,Whi00,Kyo08}

Equally appealing are the rheological properties of colloidal suspensions, complex viscoelastic fluids where the individual particles can be directly observed
by optical means, and their interactions tuned by external fields.~\cite{Bes09}
As such, confocal microscopy of sheared bulk samples has revealed rich dynamics, including the emergence of thinning,~\cite{Cheng11} thickening~\cite{Lin15}
or shear banding instabilities~\cite{Der04} in crystals~\cite{Coh06} and glasses.~\cite{Bes07}
When confined by gravity to a two-dimensional (2D) plane, an ensemble of colloids is more difficult to shear, since particles tend to escape to the bulk due to
compression or thermal fluctuations.
Thus, the use of alternative driving methods, such as magnetic fields~\cite{Yel05,Tie07} or optical tweezers,~\cite{Tho12,Zij14} may help creating compact clusters that
can be confined and deformed by the applied drive.
Recent experiments with optically confined microspheres have addressed the pressure exerted
to the boundary~\cite{Wil13} and the transmission of torque from the boundary to the center,~\cite{Wil16} while
the rich and complex rheological properties of these systems remain unexplored. In addition,
the interplay between shear and confinement gives rise to a host of new phenomena
including buckling instabilities, transport via density waves, heterogeneities
and defects that have been explored in the past theoretically~\cite{Van12,Tho12,Ger16,For16} and experimentally on 2D~\cite{Edw02,Cho11,Ivo15} and
3D colloidal systems.~\cite{Coh04,Lin09,Ram17}

In this article we investigate the dynamics and deformations of a circular colloidal cluster composed by interacting microspheres that are assembled and sheared via two independent driving fields.
Even in the absence of
shear, the particles arrange into four concentrical layers,
a generic effect in strongly confined systems.~\cite{Kla08}
Here, we use optical tweezers to assemble and rotate the outer particle layer. The other driving mechanism is a rotating magnetic field, which independently
imposes a torque on a triplet of particles (trimer) located at the center of the system.
In this 2D TC geometry, we observe velocity profiles where neighboring layers of particles slide with respect to each other creating localized shear zones.
By fixing the outer layer, we use the inner paramagnetic trimer as a microrheometer and thereby provide a realization of the original Couette experiment~\cite{Cou90} on the colloidal length scale.
Further, we observe that for a large magnetic torque, the fast spinning of the inner trimer generates a strong hydrodynamic flow and exerts a radial pressure
pushing the third layer against the outer one.
The corresponding change in the local densities of the layers manifests as a
simultaneous shear thinning of the second layer and thickening of the third one.

\section{Experimental details}
The experiments are performed using optical tweezers with an Acousto Optic Device (AOD) and a rotating magnetic field.
An infrared laser beam (ManLight ML5-CW-P/TKS-OTS, 5W maximum power, operated at 3W, $\lambda = 1064 \rm{nm}$) is deflected by the AOD (AA Optoelectronics DTSXY-400-1064, driven by a RF generator DDSPA2X-D431b-34) and then focused from above by a microscope objective (Nikon 40x CFI APO) into a closed chamber of $\sim100\rm{\mu m}$ thickness filled with the colloidal suspension.
The bottom of the chamber is observed from the other side by a second microscope objective (Nikon 40x Plan Fluor) which projects an image onto a CMOS Camera (Ximea MQ003MG-CM).
The deflection angle of the AOD is thus mapped to the position of the optical tweezers in the observation plane. The colloidal particles are trapped in the transverse direction by the gradient force of the beam and in the axial direction by the balance between the scattering force and the electrostatic repulsion with the glass substrate.

We generate a sequence of 21 optical traps equally spaced on a circle with a radius of $14.1\;\rm{\mu m}$.
Since the AOD system is capable of deflecting the laser beam every $20\rm{\mu s}$, each trap is visited once every $420\rm{\mu s}$.
The optical potential thus created can be considered as quasi-static, since the beam scanning is much faster than the typical self-diffusion time of the particles, $\tau_D\sim 40 \rm{s}$.
We then use custom made software to simultaneously rotate all the traps at different rates ranging from 0.1 to 0.6 $\rm{rad\,s^{-1}}$.

We use a bidisperse solution of paramagnetic microspheres (Dynabeads M-450 Epoxy, $4.5\rm{\mu m}$ diameter) and polystyrene particles (Invitrogen, $4\rm{\mu m}$  diameter).
One milliliter of the stock solution of the paramagnetic colloids is first dispersed in $100\rm{ml}$ of water with sodium dodecyl sulfate (SDS) solution ($\sim 0.9$ of the CMC) and the resulting mixture is stirred for two hours.
We then mix $100\rm{\mu l}$ of this solution with $1\rm{ml}$ of a mixture of sodium hydroxide and water (10mg per liter) to neutralize the SDS, and $15\rm{\mu l}$ of polystyrene particles. A drop of the final solution is placed on a capillary chamber that
was previously made by sandwiching a layer of parafilm between two coverglasses and pressing while it is heated.
The chamber is then sealed with a two component epoxy glue.

\begin{figure}[t]
  \begin{center}
    \includegraphics[width=\columnwidth,keepaspectratio]{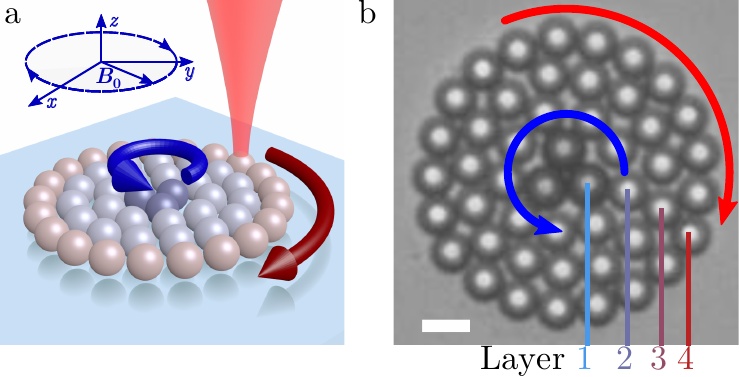}
\caption{
(a) Scheme of the experimental system where a cluster of $48$ microspheres is confined by $21$ time-shared optical traps.
The three inner particles are paramagnetic colloids that
are subjected to an in-plane rotating magnetic field of amplitude $B_0$.
Red (blue) arrow indicates the rotation direction induced by the optical (magnetic) field.
(b) Microscope image of one colloidal cluster.
Scale bar is $5\, \rm {\mu m}$, see also VideoS1 in the Supporting Information.
}
\label{fig_1}
  \end{center}
\end{figure}

The density mismatch between water and the colloidal particles makes them sediment toward the chamber bottom.
We then use a single laser trap to get three magnetic colloids together.
Afterwards, we use the tweezers to build the system from inside out, until we finally have $3$ magnetic colloids in the center, $9$ in the second layer, $15$ in the third layer and $21$ in the last layer.
The iron oxide doping of the paramagnetic particles makes them slightly darker under brightfield microscopy, which allows us to distinguish them from the other particles when assembling the cluster.
Finally, only the 21 particles in the outer layer are trapped.

\begin{figure*}[t]
  \begin{center}
    \includegraphics[width=\textwidth,keepaspectratio]{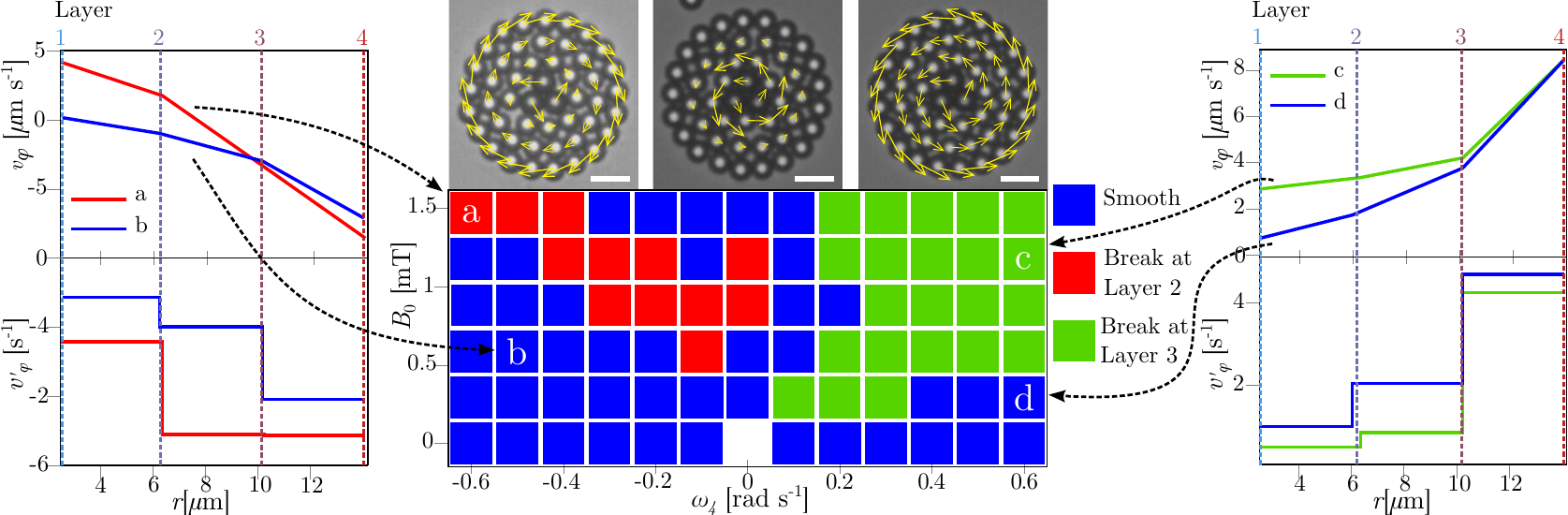}
    \caption{
      Left and right columns:
      azimuthal velocity $v_{\varphi}$ of the different layers (top) and first derivative $\dv{v}_{\varphi}$ (bottom) versus distance $r$ from the center of the cluster.
      Central column, top:
      microscope images of the colloidal cluster with the flow field superimposed as yellow arrows, see videos in the Supporting Information.
      Scale bar is $10 \rm{\mu m}$ for all images.
      Bottom: flow diagram in the $(\omega_4,B_0)$ plane illustrating the dynamic regimes observed, with smooth velocity profile (blue), a localized break at $r=6\rm{\mu m}$  (II particle layer, red) or at $r=10\rm{\mu m}$ (III particle layer green).
Here $B_0$ denotes the amplitude of the rotating field that applies a torque to the inner trimer, and $\omega_4$ the angular velocity of the optically-driven outer layer.}
    \label{fig_2}
  \end{center}
\end{figure*}
We apply a rotating magnetic field using two pairs of coils whose axes are perpendicular to each other and to the vector normal to the surface of the coverglass.
The coils are powered by two amplifiers (KEPCO BOP 20-10M) connected to a digital-to-analogue card (NI 9263).
The applied in-plane rotating field of amplitude
$B_0$ and driving frequency $\Omega$ .
During all experiments, we keep fixed the angular frequency to  $\Omega=125.7 \;\rm{rad \, s^{-1}}$, and vary $B_0$ in order to control the rotational speed of the inner triplet.

\section{The colloidal Taylor-Couette Geometry}
The system geometry is shown schematically in Fig.1(a) for the counter-rotating case, see also VideoS1 in the Supporting Infomation.
We assemble a circular cluster composed of $48$ microspheres
and radius $R=14.1 \rm{\mu m}$, by trapping the outer $21$ particles with the infrared laser.
since the AOD scanning is much faster than the
typical self-diffusion time of the particles, $\tau_D=40\rm{s}$,
the optical traps can then be considered as 21 independent harmonic potentials placed along the circle.
The outer colloidal layer is either rotated with a constant angular velocity $\omega_4 \in \pm[0.1, 0.6] \rm{rad s^{-1}}$
or P\'eclet number $\rm{Pe} \in [1.7,10.1]$
when used
in the TC geometry, or kept fixed ($\omega_4 =0\;\rm{rad s^{-1}}$) when using the inner trimer as a microrheometer.
Here we define $\rm{Pe}=\tau_D/\tau_{\omega}$ where
$\tau_{\omega}$ is the time required by a driven particle to 
travel its diameter
(here $\tau_{\omega}\sim 1/\omega$ for the trimer and $\tau_{\omega}\sim 0.28/\omega$ for the outer shell).
The internal trimer is rotated by subjecting
the paramagnetic colloids to a rotating magnetic field with
amplitude $B_0$ and angular velocity $\Omega$, ${\bm B}(t) \equiv B_0 [\cos{(\Omega t)}\basis{x}- \sin{(\Omega t)}\basis{y}]$.
The applied modulation induces the assembly of the paramagnetic particles due to
time-averaged attractive magnetic interactions,~\cite{note0} and also induces a finite torque ${\bm T}_m\sim B_0^2$ that forces the trimer to rotate at an angular velocity ${\omega_1}$.
This torque results from the internal relaxation of the particle magnetization~\cite{Jan09,Ceb11} and can be calculated
for the whole trimer as ${\bm T}_m= V_c \bm{M}\times \bm{B} = V_c B_0^2 \chi_{eff}^{"}(\Omega)\basis{z}/\mu_0$,
where $\mu_0 = 4 \pi \cdot 10^{-7}  H \, m^{-1}$, $V_c$ is the volume of the trimer, and $\chi_{eff}^{''} = 0.18$ is the effective dynamic magnetic susceptibility. Thus, at a constant driving frequency of $\Omega=20 \pi \rm{rad s^{-1}}$, ${\bm T}_m\sim B_0^2$, the amplitude of the rotating magnetic field is used to vary the rotational motion of the inner trimer.
Using video-microscopy, we measure the polar coordinates ($r_i,\varphi_i$) of each particle $i$ with respect to the center of the cluster.
We then obtain the average angular velocity per layer $n$ as $\omega_{n} = \qd{N_n^{-1}\sum_i^{N_n} \omega\lr{\bv{r}_i} }$, where the
summation counts only the $N_n$ particles within the respective layer,  Fig.1(b). From these data,
we calculate the azimuthal flow velocity, $v_{\varphi}(r)=\qd{\omega(r)}\,r$ and the first derivative, 
$dv_{\varphi}/dr = \dv{v}_{\varphi}$
in order to visualize the flow discontinuities.

\begin{figure*}[t]
\begin{center}
\includegraphics[width=\textwidth,keepaspectratio]{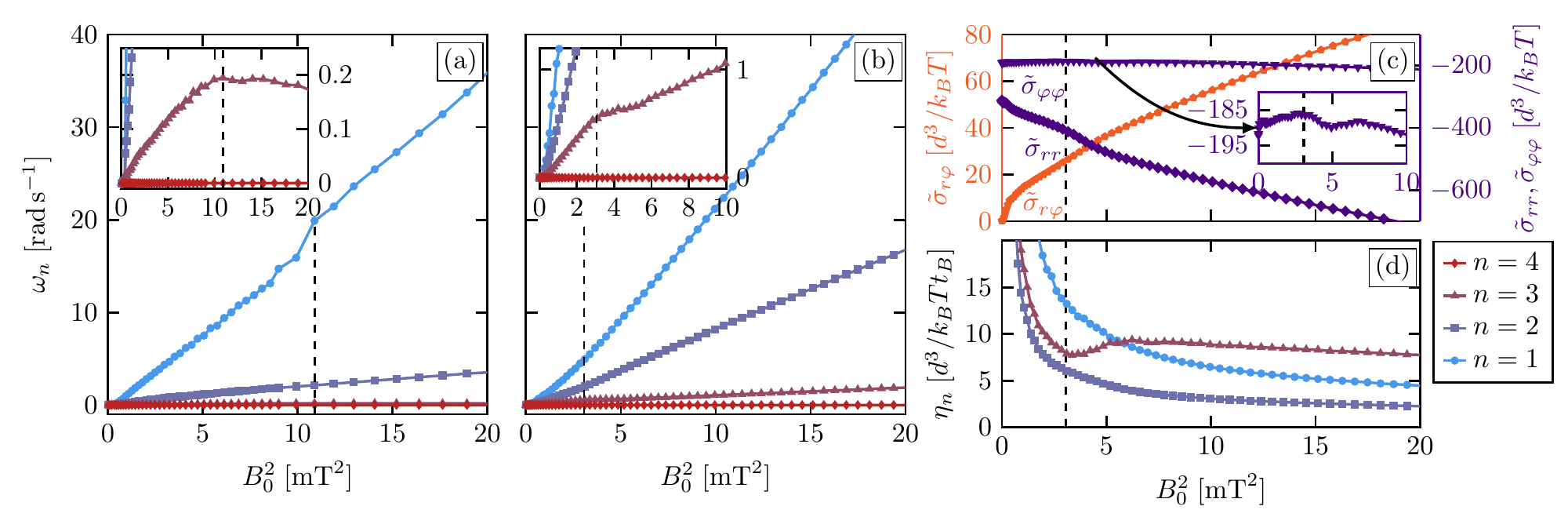}
\caption{
(a,b) Time averaged angular velocities $\omega_n$ for different layers $n$ versus squared amplitude of the applied field $B_0^2$ obtained
from experiments (a) and Brownian dynamics (BD) simulations (b). Here the magnetic torque ${\bm T}_m\sim B_0^2$.
In both cases, the inset illustrates enlargements of the main graphs.
(c) The three components of the system-averaged stress tensor, $\tilde{\sigma}_{r \varphi}=\tilde{\sigma}_{\varphi r}$ (orange), $\tilde{\sigma}_{r r}$ and $\tilde{\sigma}_{\varphi\varphi}$ (purple)
versus $B_0^2$ from BD simulations, the inset shows a zoom on $\tilde{\sigma}_{\varphi\varphi}$. (d) Viscosity $\eta_n$ of
the different layers $n$ [see \eref{shear viscosity per layer}] versus $B_0^2$ from BD simulations. The dashed black line indicates the threshold fields $B_c^2$.}
\label{fig_3}
\end{center}
\end{figure*}

In the central panel of Fig. 2 we show the complete flow diagram of our system obtained by varying the angular velocity $\omega_4$ of the outer shell and the amplitude $B_0$ of the applied magnetic field.
We classify the different
dynamical phases in terms of the velocity profiles and corresponding first derivatives, as these quantities vary along the radial direction.
Due to the strong confinement of our colloidal cluster, we never observed particle exchange between the layers.
Thus, our system does not display swirls and large scale rearrangements as observed in other sheared granular systems~\cite{How99,Mue00,Fen04,Beh08}.
In contrast to such works,
our confined system allows investigating
commensurability effects between different layers.
These layers generate periodic corrugations that slide past each other during their relative motion, and the strong confinement favors the fitting of particles of one layer in the interstices generated by the colloids of the neighboring layers.
Further, size difference between the magnetic and non magnetic particles
and the circular confinement frustrates ordering.
Thus, depending on the directions and amplitudes of the shearing torques, the system may show layers that slip and others that temporary lock to each other.
In order to characterize the different dynamical regimes,
we plot the time-averaged azimuthal velocity per layer
and the corresponding  jump in the first derivative.
Some examples are shown in the left and right panels of Fig.2.
In particular, we classify as "smooth" (blue regions in Fig.2) velocity profiles
that lead to a jump in the derivative of the azimuthal velocity
smaller than a given threshold,
in this case $|(\dv{v}_{\varphi}\lr{r_n}-\dv{v}_{\varphi}\lr{r_{n+1}})/(\rm{min}_{\dv{v}_{\varphi}}-\rm{max}_{\dv{v}_{\varphi}})|\leq 0.15$.
On the contrary, in the red and orange regions the system displays
velocity profiles
with a strong discontinuity,
leading to a pronounced jump ($> 0.15$) in $\dv{v}_{\varphi}$.
This shear-induced breakage is usually observed when there is no dominant driving mechanism, as opposed to when either the driving of the inner or outer layer is much stronger than the other.
The breakage can be localized either between the first and the second layer ($r=6 \rm{\mu m}$) or the second and third layer ($r=10 \rm{\mu m}$).
The first case occurs in the counter-rotating situation, when the inner trimer is able to drag the second layer of particles in opposite
direction than the outer layer, and the cluster inevitably breaks in pairs of counter-propagating colloidal domains (VideoS1 in the Supporting Information).
In the co-rotating case, the breakage rather localizes close to the outer layer, as this colloidal shell has a stronger capability to drag nearest  layers.
The third layer is mainly driven by the fourth one, while the second has a reduced angular speed as it also tries to follow the rotating magnetic triplet.
When the first and fourth layer have the same angular velocity
($\omega_1 = \omega_4$), the system displays a Poiseuille-like flow profile.
For $\omega_4=0$ we observe a re-entrant behavior by increasing $B_0$. For $B_0<0.75 \rm{mT}$ the trimer cannot drag the second particle layer, and the velocity profile is smooth. For $0.7\rm{mT}<B_0<1.3\rm{mT}$, the trimer drags only the second layer and there is a jump in $\dv{v}_{\varphi}$, while for $B_0>1.3\rm{mT}$ also the third layer is mobilized and the velocity profile becomes smooth again.

\section{The colloidal micro-rheometer}
We will now focus on the central region of the flow diagram in Fig.2, where we keep the outer layer fixed ($\omega_4=0$) and continuously rotate only the inner trimer within the range $\omega_1\in[0.05,36.4] \rm{rad s^{-1}}$ or $\rm{Pe} \in [0.4,2184]$.
In Fig.3(a) we show experimental measurements of the average angular velocities of the four colloidal layers versus the square of the magnetic field up to $B_0^2=20 \rm{mT}^2$.
Above a depinning threshold $B_0=0.6 \rm{mT}$, below which all particles
are at rest, the trimer starts rotating and shows an angular velocity that increases linearly with $B_0^2$, up to a threshold field strength $B_c=3.3 \, {\rm mT}$.
At $B_c$ we can see a sharp jump in the angular velocity of the trimer, where the mobility, i.e. slope of $\omega_1$, remains constant.
Further, we observe that
the slope of $\omega_3$ reveals an abrupt decrease as the magnetic torque applied to the trimer
increases, inset in Fig.3(a).
As we will show later, this
behavior is related to
a transition from "thinning" to "thickening" at $B_c$.
Above $B_c$ the fast spinning of the trimer generates a strong hydrodynamic flow that lubricates the region between the first and the second layer,
see also VideoS5 in the Supporting Information.
Thus, we find an overall thinning
of the trimer viscosity for $B_0 > B_c$.
This flow also pushes the third layer of particles
towards the outer one, increasing the local packing density and reducing the effective mobility of this layer. This leads to
an abrupt transition to thickening as seen in the third layer at a critical field
strength $B_c$.

In order to access the viscosity of all the layers and the system shear stresses, we perform $2D$ Brownian dynamics simulations of a
binary mixture of charged colloids partially confined by harmonic potentials and paramagnetic colloids, see Appendix A for more details.
Taking into account hydrodynamic effects on the Rotne-Prager level, the overdamped equation of motion for the position $\bm{r}_i$ of each colloid $i$ is given by
\begin{equation}
\label{eq:equation of motion}
\frac{d\bm{r}_i}{dt} = \sum_j \lr{\mat{\mu}^{\text{TT}}_{ij} \bv{F}_j + \mat{\mu}^{\text{TR}}_{ij} \bv{T}_j} + \Gamma_i\lr{t}\text{,}
\end{equation}
where $\mat{\mu}^{TT}_{ij}$ and $\mat{\mu}^{TR}_{ij}$ are the mobility matrices due to translation-translation (TT) and translation-rotation (TR) couplings.
The total force on the particle $j$ is given by $\bv{F}_j=\sum_{k \neq j}\bv{F}_{pp}(\bv{r}_{kj})+\bv{F}_{T}(\bv{r}_{j},t)$, and is due to the particle-particle interaction ($\bv{F}_{pp}$)
and the optical traps ($\bv{F}_{T}$), see Appendix A for more details.
The torque acting on the paramagnetic particles is given by $\bv{T}_j$, $\Gamma_i$ is a random
force, stemming from random
displacements with zero mean and variance $2D_0 \delta t$, and $D_0 \approx 0.4\rm{\mu m^2 / s}$ is the experimentally measured diffusion constant.
The simulation time scale is set to the self-diffusion time $\tau_D = d^2/D_0 \approx 40 \rm{s}$, while the discrete time step is $\delta t = 10^{-6} \tau_D$.

The results of our theoretical model are shown in Fig.3(b), and placed on the same axis as the experimental data in Fig.3(a).
We find that the model allows to qualitatively capture the dynamic features observed in the experiments.
From the simulation results, the threshold field strength is observed earlier at $B_c= 1.96 \; \rm{mT}$, which is of the same order of magnitude as in the experiment.
Moreover, the simulations allow us to access all the rheological quantities of interest such as the components of the system averaged stress tensor, $\tilde{\sigma}_{r r}$, $\tilde{\sigma}_{\varphi \varphi}$ and $\tilde{\sigma}_{\varphi r}=\tilde{\sigma}_{r \varphi}$, as well as the shear viscosity of each layer, $\eta_n$.
The former are calculated by using the virial expression for the stress tensor in polar coordinates [see \eref{stress tensor} in the Appendix].
Given the \emph{system-averaged} shear stress $\tilde{\sigma}_{r\varphi}$, we can approximate the \emph{local} shear stress by $\sigma_{r\varphi}\lr{r} \propto \tilde{\sigma}_{r\varphi}/r^2$.
This relation follows from the fact that the divergence of the stress tensor in a steady state vanishes and accounts for the spatial dependence of the volume elements in polar coordinates.
Note that this relation is exact for an incompressible Newtonian fluid.
We then approximate the local shear viscosity by the ratio
\begin{equation}\eqLabel{shear viscosity per layer}
\eta_n = \sigma_{r\varphi}\lr{r_n}/\Delta\omega_n \text{,}
\end{equation}
where $\Delta \omega_n = \omega_n - \omega_4$ is the angular velocity difference between the $n$th and the fourth layer, which is kept static, i.e. $\omega_4 = 0$.
For details see Appendix A.

\begin{figure}[t]
  \begin{center}
  \includegraphics[width=\columnwidth,keepaspectratio]{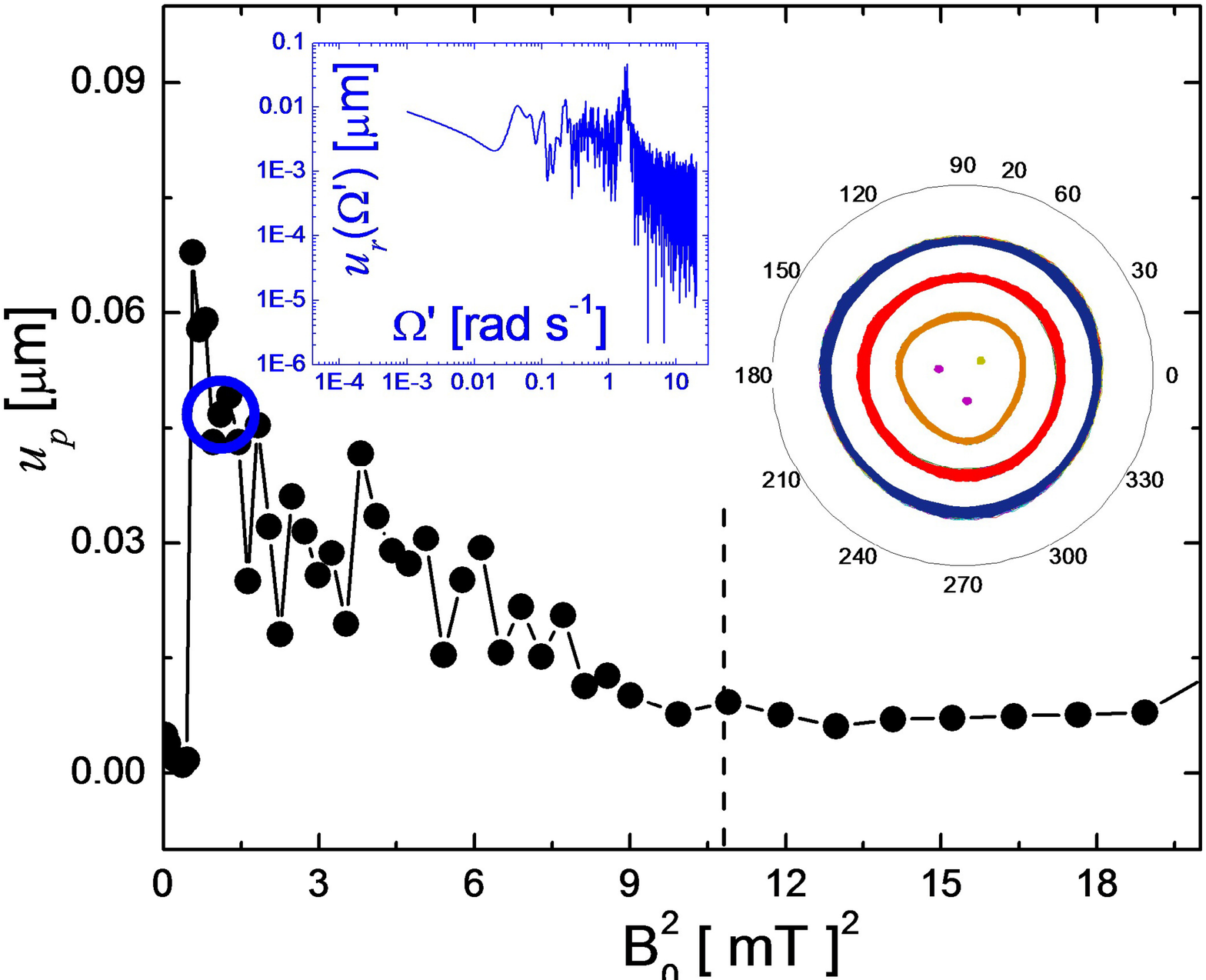}
  \caption{
    Amplitude of the radial deformation $u_{p}$ versus square of the magnetic field $B_0^2$ for a fixed outer layer of particles ($\omega_4=0$).
    Left inset: power spectrum of the elastic deformations measured for $B_0^2= 1.1\;\rm{mT^2}$.
    Right  inset: corresponding polar plot in the reference frame of the trimer of the particle trajectories showing the threefold tidal wave.}
  \label{fig_4}
  \end{center}
\end{figure}
The shear viscosities of each inner colloidal layer are plotted as a function of $B_0^2$ in Fig.3(d), and illustrate the continuous thinning of the first three layers as well as the thickening of the third one above $B_c$.
We note that, being fixed by the optical tweezers ($\omega_4=0$),
the viscosity, as defined above, of the outer layer diverges and thus it is not reported in Fig.3(d).
The same holds for the inner layers for $B_0\rightarrow 0$, being pinned to the outer one.
The transition at $B_c$ is also reflected in all components of the stress tensors shown in Fig.3(c).
In particular, the shear stress ($\tilde{\sigma}_{r \varphi}$) displays an increase in the slope at $B_c^2$, and subsequent decrease at high  magnetic field strengths, corresponding to the shear
thinning and thickening respectively.
The two diagonal components of the stress tensor decrease as the field increases, corresponding to an increase of the radial $P_r = -\tilde{\sigma}_{r r}$ and azimuthal pressure $P_\varphi = -\tilde{\sigma}_{\varphi\varphi}$.
The shear thickening at $B_c$ is accompanied by a steep increase of $P_r$, which is resolved subsequently.
Further, we have confirmed this general behavior by performing simulations on a larger cluster with $N=76$ particles. This system displays the same type of local thinning and thickening in the layer closest to the fixed outer layer, even though the associated field is larger.

\section{Radial deformations}
While the dynamic transition at $B_c$ appears to be sharp in terms of the different rheological quantities, we find in the experiments that the cluster shows a continuous
decrease in the amplitude of the radial distortions induced by the rotating trimer.
Given the non-circular shape of the trimer, the particles from the second layer are forced to periodically enter and exit its interstitial regions.
The induced deformation thus appears in form of a threefold tidal wave as shown in the small polar plot in Fig.4.
To characterize these elastic deformations, we measure their amplitude as a function of the magnetic field strength in the Fourier space and in the reference frame of the trimer as,
$u_{r}(\Omega')=N_2^{-1}\sum_{j=1}^{N_2}T^{-1}\int_0^T\exp{[3\mathrm{i}(\Omega't-\varphi_j)]}r(\varphi_j,t)dt$,
where $r(\varphi_j,t)$ is the distance of the $N_2$
particles composing the second layer and located at an angle $\varphi_j$ in the reference frame of the trimer.
In Fig.4 we plot the measured peak of $u_r (\Omega')$,
namely $u_{p} = \rm{max}[u_r(\Omega'), \Omega'\in {\rm I\!R}]$,
as a function of $B_0^2$ and in the upper left inset the corresponding
$u_r (\Omega')$  for one exemplary $B_0$.
For small field strengths ($B_0^2 < B_c^2$), the radial path of the particles in the second layer is deformed by the rotating trimer.
After a transient regime, the amplitude of the deformation stabilizes to a stationary value where the relative speed between the $N_2$ composing particles approaches zero.
We further note that, being in the Stokes regime, the decoupling of the dynamics along the radial and azimuthal
directions is effectively possible, and
this explains the emergence of two different types of dynamic transitions
(of sharp and continuous character, respectively) along the two directions.

\section{Conclusions}
We have realized a colloidal microrheometer based on the combined action of magnetic and optical torques.
The present approach may be used as an effective microrheological tool to explore the viscoelastic properties of complex fluids.
This could potentially include biological media confined between the magnetic trimer and the optically trapped colloidal ring.
Further, by replacing the internal trimer with a single ferromagnetic particle, one could simplify the system geometry, decoupling the information measured along the shear and radial direction.
In addition, the optical tweezers may be programmed to periodically shear the outer layer of particles and thus to explore the
frequency-dependent properties of complex fluids, which represents an exciting future avenue.

\section*{Appendix}
\appendix

\section{Model and simulation details}\label{sec:Appendix A}
Here, we describe the microscopic model underlying our Brownian dynamics simulations.
We consider a dense binary mixture of charged colloids and paramagnetic colloids confined to a 2D disk-like area by an outer layer of particles trapped by harmonic potentials.
All colloids interact via a repulsive Yukawa potential
\begin{equation}\label{eq:particle interaction}
  U_{pp}\lr{r_{ij}} = V_0 \exp\LR{\kappa\lr{d_{ij}-d}} \frac{ \exp\lr{-\kappa r_{ij} } }{ \kappa r_{ij} }\text{,}
\end{equation}
with $V_0$ being the particle-particle interaction strength, $\kappa$ the inverse Debye screening length, $d_{ij} = (d_i+d_j)/2$ the mean particle diameter, and $r_{ij} = \left| \bv{r}_i - \bv{r}_j \right|$ the distance between particle $i$ and $j$.
To model the repulsive particle interactions corresponding to the experimental system, we set $\kappa = 70 d^{-1}$ and $V_0 = 1.69037\cdot10^{33} k_B T$.

The paramagnetic particles interact, in addition to the repulsion,
via the time-averaged dipole-dipole (DD) interaction \cite{Jaeger2013, Martinez-Pedrero2015}
\begin{equation}\label{eq:dipole dipole interaction}
U_\text{DD} \lr{ r_{ij} } = -\frac{ \lr{ V \chi B_{0,\text{DD} } }^2 }{8 \pi \mu_0 r_{ij}^3 }\text{,}
\end{equation}
with $\mu_0 = 4\pi \, 10^{-7} \,Hm^{-1}$ being the magnetic permittivity, $V = (4\pi/3) d_i^3 $ the particle volume, $\chi=1.4$ the effective volume susceptibility, and $B_{0,\text{DD} }$ the amplitude of the magnetic field used for the DD interactions
We note that in the experiments $B_{0,\text{DD}}$ and $B_0$ are expected to be equal.
However, in our simulations this leads to an unintentional compression of the inner layer, due to the finite softness of the particle interactions.
To counteract this artifact, the amplitude of the rotating magnetic field determining the strength of the DD interactions is fixed to a constant value (specifically, here we set $B_{0,\text{DD}} = 700\;\rm{Am^{-1}}$).

The particles forming the outer colloidal layer are trapped by harmonic potentials
\begin{equation}\label{eq:ring potential}
  U_{R}\lr{\bv{r}_i, t} = \frac{k}{2} \left| \bv{r}_i - \bv{r}_{i,0}\lr{t} \right|^2 \quad\quad i\in\text{outer layer}\text{,}
\end{equation}
where $i$ is from the $N_4$ particles of the outer layer, $k$ is the strength of the harmonic potential, and $\bv{r}_{i,0}$ is the position of the center of the trap corresponding to the particle $i$.
The constant angular velocity of the outer layer is applied by translating the center of the harmonic traps on a ring, where the corresponding trajectories are given by
\begin{align}
\label{eq:ring position}
  \bv{r}_{j,0} \lr{t} &= R \LR{ \cos\lr{ \Phi_{j}\lr{t} } \bv{e}_x + \sin\lr{ \Phi_{j}\lr{t} } \bv{e}_y } \\
  \Phi_{j}\lr{t} &= 2\pi\lr{\frac{j}{N_4} + \omega_4 t}\text{,}
\end{align}
with $R$ being the radius of the outer layer, and $\omega_4$ its angular velocity.

Following earlier simulation studies of driven colloids,\cite{Gauger2009, Hansen2011} we model the hydrodynamic interactions on the Rotne-Prager level.
Furthermore, to account for the hydrodynamic effect of the planar substrate, i.e a liquid-solid interface, we apply the Blake's solutions.~\cite{Blake1971, Gauger2009, Hansen2011}
The resulting hydrodynamic couplings are described by translation-translation (TT) and translation-rotation (TR) mobility matrices entering the equations of motion [see Eq.~(1) in the main text].
Specifically, the TT mobility matrix is given by:
\begin{equation}\label{eq:trans-trans mobility}
  \mat{\mu}_{ij}^{TT} = \frac{D_0}{k_B T} \mat{I} + \mat{G}^{RP}\lr{ \bv{r}_{ij} } - \mat{G}^{RP}\lr{ \bv{R}_{ij} } + \delta \mat{G}\lr{ \bv{R}_{ij} } \text{,}
\end{equation}
where $\bv{r}_{ij} = \bv{r}_i - \bv{r}_j$ is the vector between particle $i$ and $j$, $\bv{R}_{ij} = \bv{r}_i - \bv{r}'_j$ is the vector between particle $i$ and the image of particle $j$, $\bv{r}'_j = (x_j,y_j,-z_j)^T$ is the position of the image of particle $j$ for an interface which is located at $z=0$ and is infinitely extended in $x$ and $y$ direction.
The first term of \eref{trans-trans mobility} corresponds to the self mobility of particle $i$.
The second and third term are the Rotne-Prager tensors for the TT coupling of
particle $i$ to particle $j$ and its image respectively, given by
\begin{equation*}\label{eq:rotne-prager}
  \mat{G}^{RP}\lr{ \bv{r}_{ij} } = \frac{1}{3 \pi \eta d_i} \LR{  \frac{3}{8} \frac{d_j}{r_{ij}}\lr{ \mat{I} + \bv{r}_{ij} \otimes \bv{r}_{ij}  } + \frac{1}{16}\lr{\frac{d_j}{r_{ij} } }^3 \lr{ \mat{I} - 3 \bv{r}_{ij} \otimes \bv{r}_{ij} }  }\text{,}
\end{equation*}
where $\eta = k_B T / 3\pi d_j D_0$ is the effective viscosity of particle $j$ given by the Stokes-Einstein relation.
The last term in \eref{trans-trans mobility} contains the correction terms according to the Blake solution (on the Rotne-Prager level)
%
\begin{align}\label{eq:cross corrections}
  \delta G_{\alpha\alpha}\lr{\bv{R}_{ij}} &= \frac{1}{3 \pi \eta d_i} \LR{ \frac{3}{16}d_j R_z^2 \lr{ \frac{1}{R_{ij}^3} - 3\frac{R_\alpha^2}{R_{ij}^5} } - \frac{d_j^5}{64}\frac{1}{R_{ij}^9}\lr{4 R_\alpha^4 - R_\beta^4 + 3 R_\alpha^2 R_\beta^2} } \\
  \delta G_{\alpha\beta}\lr{\bv{R}_{ij}} &= \frac{1}{3 \pi \eta d_i} \LR{ \frac{9}{4}d_j R_z^2 \frac{R_\alpha^2}{R_{ij}^5} - \frac{d_j^5}{64}\lr{R_\alpha^2 + R_\beta^2}\frac{R_\alpha R_\beta}{R_{ij}^9} }\text{,}
\end{align}
%
where $\alpha,\beta \in \left\{x,y\right\}$ and $R_\alpha$, $R_\beta$ are the $\alpha$
and $\beta$ components of $\bv{R}_{ij}$, respectively.\\
The paramagnetic particles, forming the inner layer $n=1$, are driven by the rotating magnetic field which exerts a constant torque onto these particles,
\begin{equation}\label{eq:torque}
  \bv{T}_j = \frac{ V \chi \tilde{B}_0^2 t_\text{rel} \omega_m }{\mu_0 (1 + t_\text{rel}^2 \omega_m^2) } \basis{z}\text{,}
\end{equation}
where $\tilde{B}_0$ is the magnetic field strength employed in the
simulations \footnote{
The magnetic field strength in the simulations is set to $\tilde{B}_0 = \beta B_0\sqrt{\tau_D}$, where the factor of proportionality $\beta\approx 862$ is determined by fitting the linear response of the free trimer, data not shown here.
}, $t_{rel} = 10^{-4} \rm{s}$ is the relaxation time and $\omega_m = 2\pi \;\rm{rad\,s^{-1}}$ is the angular velocity of the magnetic field.
This torque is transferred to the motion of particles via the TR mobility matrix
\begin{equation}\label{eq:rotation translation mobility}
  \mat{\mu}^{TR}_{ij} = \frac{1}{8\pi\eta}\lr{ \frac{ \bv{r}_{ij}}{r^3_{ij}} - \frac{ \bv{R}_{ij}}{R^3_{ij}} }\mat{\hat{\epsilon}}\text{,}
\end{equation}
where $\mat{\hat{\epsilon}}$ is the Levi-Civita symbol.\\
The inner particles of the colloidal cluster are then driven by a flow field $\bv{u}_i^{TR} = \sum_j\mat{\mu}^{TR}_{ij} \bv{T}_j$ exerted from the individually rotating paramagnetic colloids.
In the absence of all other particles, the cluster of the three paramagnetic particles performs a regular circular motion on a ring, with an angular velocity proportional to $B_0^2$.

Combining the hydrodynamic mobility matrices with the equation of motion of conventional (overdamped) Brownian dynamics simulations, we finally arrive at Eq.(1) in the manuscript.
Following previous studies,~\cite{Frenkel2001} the interaction potentials were truncated and shifted accordingly at $r_{c,pp} = 1.169d$ for the Yukawa interaction [see \eref{particle interaction}] and at $r_{c,DD} = 10.775d$ for the DD interaction [see \eref{dipole dipole interaction}], with $d \approx 4\;\rm{\mu m}$ being the diameter of the non-paramagnetic particles.
The diameter of the paramagnetic particles is set to $d_{inner} = 1.125d$.

From the particle positions, we calculate the mean angular velocity per layer as
\begin{equation}
\omega_n=\qd{\frac{1}{2\pi N_n}\sum_{i=1}^{N_n}\frac{\varphi_i(t+dt)-\varphi_i(t)}{dt}} \; ,
\end{equation}
with $N_n$ the number of particles in the layer $n$,
$dt$ the time interval between the measurements, and $\varphi_i$ the angular component in polar coordinates of particle $i$, which is defined as
\begin{equation}
\bm{r}_i=r_i [\cos{(\varphi_i)}\basis{x}+\sin{(\varphi_i)}\basis{y}] \, \, .
\end{equation}
Further, we calculate the elements of the \emph{system-averaged} stress tensor in polar coordinates as
\begin{equation}\eqLabel{stress tensor}
\tilde{\sigma}_{nm}= \qd{ -\frac{1}{V} \sum_{i=1}^{N}\sum_{i\neq j}^{N}\frac{\partial U_{pp}(r_{ij})}{\partial r_{ij}}\frac{[\bm{r}_{ij}\cdot \basis{n}(\varphi_i)][\bm{r}_{ij}\cdot \basis{m}(\varphi_i)]}{r_{ij}} }\; ,
\end{equation}
where $V = \pi R^2$ is the volume of the system, $N$ is the number of particles and $\basis{n}$, $\basis{m}$ are the orthogonal basis vectors in polar coordinates.
In the steady state of an incompressible fluid, this system-averaged stress fully determines the local stresses.
This is due to the fact that the stress density needs to be constant, i.e. the divergence of the stress tensor vanishes.
In polar coordinates the local shear stress is $\sigma_{r\varphi}\lr{r} = \alpha/r^2$, where $\alpha$ is an integration constant.
The latter can be determined by integrating the local stress over the whole system
\begin{equation}\eqLabel{local shear stress factor}
  \tilde{\sigma}_{r\varphi} := \frac{1}{\pi\lr{ r_4^2 - r_1^2} } \int_{r_1}^{r_4} \int_{0}^{2\pi} \sigma_{r\varphi}\lr{r} r \,d\varphi dr = \frac{2\log\lr{\frac{r_4}{r_1} }}{ r_4^2 - r_1^2 } \; \alpha\text{,}
\end{equation}
where $r_4 = R$ is the radius of the outer ring and $r_1$ is the radius of the paramagnetic ring.
From \eref{local shear stress factor} it follows that the local shear stress is given by
\begin{equation}\eqLabel{local shear stress}
  \sigma_{r\varphi}\lr{r} = \frac{ R^2 - r_1^2 }{2\log\lr{\frac{R}{r_1} }}  \;\frac{\tilde{\sigma}_{r\varphi}}{r^2}\text{.}
\end{equation}
Using this expression we approximate the shear viscosity per layer as the ratio between the local shear stress and the angular velocity difference
\begin{equation}\eqLabel{shear viscosity}
\eta_n=\frac{ \sigma_{\rho \varphi}\lr{r_n} }{\Delta\omega_n}\; \text{,}
\end{equation}
where $\Delta \omega_n = \omega_n - \omega_4$ is the angular velocity of layer $n$ relative to the outer ring $\omega_4$.
In writing \eref{shear viscosity} we have assumed $\Delta\omega_n$ to be proportional to the true local shear rate.
In other words, for a solid-body rotation ($\omega_n = \omega \forall n$) the shear rate would vanish, as one would expect.

\section{Acknowledgments}
We thank Thomas M. Fischer, Benjamin Dollet and Hartmut L\"owen for many stimulating discussions.
This research was funded by the ERC starting Grant "DynaMO" (No. 335040).
A.O.A. acknowledges support from the "Juan de la Cierva" program (FJCI-2015-25787).
J.O. and P.T. acknowledge support from MINECO (FIS2016-78507-C2) and DURSI (2017SGR1061).
S.G. and S. H. L. K. acknowledge support from the Deutsche Forschungsgemeinschaft through SFB 910 (project B2).

\footnotesize{
\bibliography{Biblio} 

\begin{mcitethebibliography}{47}
\bibitem[not()]{note0}
The magnetic dipolar interaction energy between a pair of particles $(i,j)$
  with moments $m_{i},m_{j}$ and located at a distance
  $\bm{r}_{ij}=|\bm{r}_{i}-\bm{r}_{j}|$ is given by, $U_{d}=(\mu_0/ 4\pi)[
  ({\bm m}_i \cdot {\bm m}_j) /r_{ij}^3 - 3({\bm m}_i \cdot {\bm r}_{ij})({\bm
  m}_j\cdot {\bm r}_{ij}) /r_{ij}^5 ]$. This potential is attractive or
  repulsive when these moments are parallel or perpendicular to the separation
  distance $\bm{r}_{ij}$. Performing a time average of the dipolar energy
  between two equal colloids for a magnetic field rotating in the $(x,y)$
  plane, gives an attractive potential in this plane $\langle U_d \rangle
  =-\frac{\mu_0 m^2}{8 \pi (x+y)^3}$, leading to clustering, as previoulsy
  reported~\cite{Tie07}\relax
\mciteBstWouldAddEndPuncttrue
\mciteSetBstMidEndSepPunct{\mcitedefaultmidpunct}
{\mcitedefaultendpunct}{\mcitedefaultseppunct}\relax
\EndOfBibitem
\end{mcitethebibliography}
\bibliographystyle{rsc} 
}

\end{document}